\documentclass[aps,prd,preprint,nofootinbib]{revtex4}

\usepackage{graphicx}
\usepackage{epsfig}

\begin{document}

\title{Determination of the $\eta$ and $\eta'$ Mixing Angle from the
Pseudoscalar Transition Form Factors}
\author{Tao Huang$^{1}$\footnote{email:
huangtao@mail.ihep.ac.cn} and Xing-Gang Wu$^{2,3}$\footnote{email:
wuxg@itp.ac.cn}}
\address{
$^1$Institute of High Energy Physics, Chinese Academy of Sciences,
P.O.Box 918(4), Beijing 100039, P.R. China\\
$^2$Institute of Theoretical Physics, Chinese Academy of Sciences,
P.O.Box 2735, Beijing 100080, P.R. China.\\
$^3$Department of Physics, Chongqing University, Chongqing 400044,
P.R. China}


\begin{abstract}

The possible range of $\eta-\eta'$ mixing angle is determined from
the transition form factors $F_{\eta \gamma}(Q^2)$ and $F_{\eta'
\gamma}(Q^2)$ with the help of the present experimental data. For
such purpose, the quark-flavor mixing scheme is adopted and the
pseudoscalar transition form factors are calculated under the
light-cone pQCD framework, where the transverse momentum corrections
and the contributions beyond the leading Fock state have been
carefully taken into consideration. We construct a phenomenological
expression to estimate the contributions to the form factors beyond
the leading Fock state based on their asymptotic behavior at $Q^2\to
0$ and $Q^2\to\infty$. By taking the quark-flavor mixing scheme, our
results lead to $\phi= 38.0^{\circ}\pm 1.0^{\circ}\pm 2.0^{\circ}$,
where the first error coming from experimental uncertainty and the
second error coming from the uncertainties of the wavefunction
parameters. The possible intrinsic charm component in $\eta$ and
$\eta'$ is discussed and our present analysis also disfavors a large
portion of intrinsic charm component in $\eta$ and $\eta'$, e.g.
$|f^c_{\eta'}|\le 50\;{\rm MeV}$.\\


\end{abstract}

\maketitle

\section{Introduction}

The light-cone (LC) formalism \cite{lc1,lc2,lc3} provides a
convenient framework for the relativistic description of hadrons in
terms of quark and gluon degrees of freedom and for the application
of pQCD to exclusive processes. Among them, the $\eta-\eta'$ mixing
is a subject of considerable interest, which has been examined in
many investigations, see e.g. experimental ones
\cite{CELLO,CLEO,l3,tpc,kloe,kloe1,babar,pluto,kroll,bes} and
theoretical ones
\cite{old1,old2,feldmann,feldmann1,feldmann2,feldmann3,old3,huangcao,ohta}.
Some experiments have been done recently, e.g. the new KLOE value of
$R_\phi=\frac{\Gamma[\phi\to\eta'\gamma]}
{\Gamma[\phi\to\eta\gamma]}=(4.9\pm0.1_{stat}\pm0.2_{syst}) \cdot
10^{-3}$ \cite{kloe1} leads to $\phi=41.2^{\circ}\pm1.2^{\circ}$
\cite{kroll}; the BES collaboration has announced a new measured
value for $R_{J/\Psi}$, i.e.
$R_{J/\Psi}=\frac{\Gamma[J/\Psi\to\eta'\gamma]}
{\Gamma[J/\Psi\to\eta\gamma]}=4.94\pm0.40$ \cite{bes} that leads to
$ \phi=38.8^{\circ}\pm 1.2^{\circ}$. Furthermore, the newly
measurements of the form factors $F_{\eta \gamma}(Q^2)$ and
$F_{\eta' \gamma}(Q^2)$ at asymptotic region by BaBar collaboration
\cite{babar}, $Q^2F_{\eta\gamma}(Q^2)|_{Q^2=112\;{\rm
GeV}^2}=0.229\pm0.030\pm0.008$~GeV and
$Q^2F_{\eta'\gamma}(Q^2)|_{Q^2=112\;{\rm
GeV}^2}=0.251\pm0.019\pm0.008$~GeV, will provide further constraints
to the theoretical predictions.

The pseudoscalar transition form factors $F_{\eta \gamma}(Q^2)$ and
$F_{\eta' \gamma}(Q^2)$ provide a good platform to study the $\eta$
and $\eta'$ mixing effects, which have already been studied in
literature by several groups
\cite{old3,feldmann3,huangcao,huang2,jakob,caofg}. However in these
calculations, either only the leading Fock-state (e.g.
Ref.\cite{jakob}) or only the asymptotic behavior of the form
factors (e.g. Ref.\cite{caofg}) have been taken into consideration
to determine the mixing angle. As has been pointed out in
Refs.\cite{huangcao,huang2}, the mixing angle can not be reliably
determined without a proper considering of the contributions from
the non-valence quark states due to the fact that even though the
higher Fock states' contributions are power suppressed in large
$Q^2$ region, they will give sizable contributions to small and
intermediate regions. In fact, it has been pointed out in
Ref.\cite{bhl} that the leading Fock state contributes to
$F_{P\gamma}(Q^2)|_{Q^2=0}$ ($P$ stands for the pseudoscalar mesons)
only half and the remaining half should be come from the higher Fock
states. The higher Fock states' contributions in small $Q^2$ region
can not be calculated by the perturbative QCD approach due to its
nonperturbative feature. Recently, Ref.\cite{huangwu} had
constructed a phenomenological expression for the pion-photon
transition form factor to estimate the contributions beyond the
leading Fock state based on its asymptotic behavior at $Q^2\to 0$
and $Q^2\to\infty$ ($Q^2$ stands for the momentum transfer in the
process). The predicted results for $F_{\pi\gamma}(Q^2)$ there agree
well with the experimental data in the whole $Q^2$ region. In the
present paper, we will adopt this newly developed method to estimate
the higher Fock states' contributions of the form factors $F_{\eta
\gamma}(Q^2)$ and $F_{\eta' \gamma}(Q^2)$, and then to derive a
possible range for the mixing angle by comparing the predicted
results with the experimental data.

As for the $\eta-\eta'$ mixing, two mixing scheme are adopted in the
literature, i.e. the octet-singlet mixing scheme and the
quark-flavor mixing scheme. These two schemes can be related with a
proper rotation of an ideal mixing angle
($\theta_{id}=-\arctan\sqrt{2}\simeq-54.7^{\circ}$)
\cite{feldmann2,feldmann4}. A dramatic simplification can be
achieved by adopting the quark-flavor mixing scheme, especially, the
decay constants in the quark-flavor basis simply follow the pattern
of state mixing due to the OZI-rule \cite{feldmann2}. Furthermore,
by adopting the quark-flavor mixing scheme and also by carefully
dealing with the higher Fock states' contributions, a naive
discussion (at the end of Sec.II.C) shows that the value of
$Q^2F_{\eta\gamma}(Q^2)$ decreases, while the value of
$Q^2F_{\eta'\gamma}(Q^2)$ increases, with the increment of the
mixing angle $\phi$, so a possible range for $\phi$ can be derived
by comparing with the experimental data on the form factors $F_{\eta
\gamma}(Q^2)$ and $F_{\eta' \gamma}(Q^2)$. And then by adopting the
relation between the two schemes as shown in
Refs.\cite{feldmann2,feldmann4}, all the three mixing angles
$\theta_P$, $\theta_1$ and $\theta_8$ involved in the octet-singlet
scheme can be determined, where $\theta_P$ is mixing angle for the
states and $\theta_{1/8}$ are mixing angles for the decay constants
$f_1$ and $f_8$. The theoretical and phenomenological considerations
performed in Refs.\cite{feldmann,feldmann3,favor3} also favor the
quark-flavor basis. Therefore, we will adopt the quark-flavor mixing
scheme to do our calculation through out the paper.

The paper is organized as follows. In Sec.II, we outline our
techniques for determining the $\eta-\eta'$ mixing angle, where
expressions of the pseudoscalar transition form factors beyond the
leading Fock State are provided. In Sec.III, we present the
numerical results for the $\eta\gamma$ and $\eta'\gamma$ transition
form factors, and then derive a possible range for the mixing angle
$\phi$ by comparing with the present experimental data on the form
factors $F_{\eta \gamma}(Q^2)$ and $F_{\eta' \gamma}(Q^2)$. Some
discussions of the uncertainty sources for $\phi$ determination are
provided in Sec.IV. The final section is reserved for summary.

\section{$\eta-\eta'$ mixing angle and Expressions of the pseudoscalar
transition form factors beyond the leading Fock State}

\subsection{Definition in the quark-flavor basis}

In the quark-flavor basis, the two orthogonal basis states are
assumed to have the following parton composition in a Fock state
description :
\begin{equation}
|\eta_q \rangle = \Psi_{\eta_q} \, \frac{|u\bar u + d\bar
d\rangle}{\sqrt{2}} + \cdots \;,\;\; |\eta_s \rangle = \Psi_{\eta_s}
\, |s\bar s \rangle + \cdots \label{quarkflavor}
\end{equation}
where $\Psi_{\eta_i}$ ($i=q$, $s$) denote the LC wavefunctions of
the corresponding parton states, and the dots stand for higher Fock
states. The physical meson states are related to the basis
(\ref{quarkflavor}) by an orthogonal transformation
\begin{eqnarray}
\left( \matrix{|\eta\phantom{{}'}\rangle \cr |\eta'\rangle } \right)
&=& U(\phi) \, \left(\matrix{|\eta_q\rangle \cr|\eta_s\rangle }
\right) \; ,\;\;\;  U(\phi) = \left(\matrix{\cos\phi & -\sin\phi \cr
\sin\phi & \phantom{-}\cos\phi} \right) \ , \label{mixangle}
\end{eqnarray}
where $\phi$ is the mixing angle. Under such scheme, the decay
constants in the quark-flavor basis simply follow the pattern of
state mixing due to the OZI-rule \cite{feldmann2}, i.e.
\begin{eqnarray}
\left(\begin{array}{cc}f_\eta^q & f_\eta^s \cr f_{\eta'}^q &
f_{\eta'}^s\end{array} \right) &=& U(\phi) \, {\rm diag}[f_q,f_s] ,
\label{decayconstant}
\end{eqnarray}
where the two basic decay constants $f_q$ and $f_s$ are defined as
\begin{equation}
f_i = 2 \sqrt{3} \int_{k_\perp^2\leq \mu_0^2} \frac{dx \,
d^2k_\perp}{16 \pi^3} \, \Psi_{\eta_i}(x,k_\perp)
\end{equation}
with $\mu_0$ the factorization scale that is of order ${\cal
O}(1\;{\rm GeV})$.

Useful constraints to determine the $\eta-\eta'$ mixing angle can be
derived by considering the two-photon decay of $\eta$ and $\eta'$.
The decay amplitudes of $\eta_s\to\gamma\gamma$ and
$\eta_q\to\gamma\gamma$ have the similar Lorentz structure as that
of $\pi^0\to\gamma\gamma$ \cite{donoghue} \footnote{It is noted that
the higher helicity states do not have contribution to the decay
amplitude.}:
\begin{equation}
{\cal A}_{P\to \gamma_1(k_1)\gamma_2(k_2)}=\frac{\alpha}{\pi}
\frac{c_P}{f_P}\epsilon^{\mu\nu\alpha\beta}\epsilon^*_\mu(k_1)
\epsilon_\nu^*(k_2)k_{1\alpha}k_{2\beta},
\end{equation}
where the fine-structure constant $\alpha=1/137$,
$c_P=(c_s,c_q)=\left(\sqrt{2}/3,5/3\right)$ for the states
$P=(\eta_s,\eta_q)$, and $f_P$ is the corresponding decay
constant. Then the decay widths for $\eta\to\gamma\gamma$ and
$\eta'\to\gamma\gamma$ can be written as:
\begin{eqnarray}
\label{etagamma}\Gamma_{\eta\to\gamma\gamma}&=&\frac{\alpha^2
M^3_{\eta}}{64\pi^3}\left(\frac{c_q \cos\phi}{f_q}- \frac{c_s
\sin\phi}{f_s}\right)^2 \;{\rm and}\;\;
\Gamma_{\eta'\to\gamma\gamma}=\frac{\alpha^2 M^3_{\eta'}}
{64\pi^3}\left(\frac{c_q \sin\phi}{f_q}+\frac{c_s
\cos\phi}{f_s}\right)^2 ,
\end{eqnarray}
where the two photon decay widths of $\eta$ and $\eta'$ and their
masses can be found in PDG \cite{pdg}
\begin{eqnarray}
&&\Gamma_{\eta\to\gamma\gamma}=0.46\pm0.04 \;{\rm KeV},\;\;
M_{\eta}=547.30\pm0.12 \;{\rm MeV}, \nonumber\\
&&\Gamma_{\eta'\to\gamma\gamma}=4.37\pm0.25\;{\rm KeV},\;\;
M_{\eta'}=957.78\pm0.14 \;{\rm MeV}.\nonumber
\end{eqnarray}

From Eq.(\ref{etagamma}), we obtain the correlation between
$f_q/f_s$ and $\phi$ :
\begin{eqnarray}\label{decayfq}
f_q &=& \frac{c_q\alpha}{8\pi^{3/2}}\left[\sqrt{
\frac{\Gamma_{\eta\to\gamma\gamma}} {M_\eta^3}}\cos\phi+\sqrt{
\frac{\Gamma_{\eta'\to\gamma\gamma}}{M_{\eta'}^3}}
\sin\phi\right]^{-1}
\end{eqnarray}
and
\begin{eqnarray} f_s &=& \frac{c_s\alpha}{8\pi^{3/2}}\left[\sqrt{
\frac{\Gamma_{\eta'\to\gamma\gamma}}{M_{\eta'}^3}}\cos\phi-\sqrt{
\frac{\Gamma_{\eta\to\gamma\gamma}}{M_\eta^3}}\sin\phi\right]^{-1}
.\label{decayfs}
\end{eqnarray}
It shows that if knowing the range of the mixing angle $\phi$, the
ranges of the decay constants $f_q$ and $f_s$ can be determined
accordingly; and vice versa.

\subsection{A brief review of $\pi\gamma$  Transition form factor}

In order to calculate the $\eta\gamma$ and $\eta'\gamma$ transition
form factors, we first give a brief review of the $\pi\gamma$
transition form factor $F_{\pi\gamma}(Q^2)$. A comprehensive
analysis of $F_{\pi\gamma}(Q^2)$ has been given in
Ref.\cite{huangwu}, in which the transverse-momentum dependence in
both the hard scattering amplitude and the LC wavefunction and the
contributions beyond the leading Fock state have been taken into
consideration. Especially, a phenomenological expression to estimate
the contributions beyond the leading Fock state has been
constructed, which is based on the form factor's asymptotic
behaviors at $Q^2\to 0$ and $Q^2\to\infty$.

As has been pointed out in Ref.\cite{huang2} that the
transverse-momentum dependence in both hard-scattering amplitude and
the meson wavefunction should be kept to give a consistent analysis
of the form factor. The revised LC harmonic oscillator model as
suggested in Ref.\cite{hms} was employed for the LC wavefunction
$\Psi_{\pi}(x,{\bf k}_{\perp})$, which is constructed based on the
Brodsky-Huang-Lepage (BHL) prescription~\cite{bhl}. More explicitly,
the LC wavefunction of $\pi^0=\frac{1}{\sqrt{2}}|u\bar{u}
-d\bar{d}\rangle$ can be written as
\begin{equation}\label{pionwave}
\Psi_{\pi}(x,{\bf k}_{\perp}) =A_{\pi}\left[\exp\left(-\frac{{\bf
k}_{\perp}^2 +m_q^2}{8{\beta_{\pi}}^2x(1-x)}
\right)\chi^K(m_q,x,{\bf k}_{\perp})\right],
\end{equation}
with the normalization constant $A_{\pi}$, the harmonic scale
$\beta_{\pi}$ and the light quark mass $m_q$ to be determined. Since
the contribution from the higher helicity states
($\lambda_1+\lambda_2=\pm 1$) has little contribution in comparison
to the usual helicity state ($\lambda_1+\lambda_2=0$), so we only
write the explicit term for the usual helicity state. The spin-space
wavefunction $\chi^K(x,{\bf k}_{\perp})$ for the usual helicity
state of pion can be written as \cite{hms}, $ \chi^K(m_q,x,{\bf
k}_{\perp}) = m_q/\sqrt{m_q^2+k_\perp^2}$, where $k_\perp=|{\bf
k}_{\perp}|$. Furthermore, one can derive a relation between $m_q$
and $\beta_{\pi}$ by adopting the constraints from $\pi^0\to\mu\nu$
and $\pi^0\to\gamma\gamma$ \cite{huangwu}
\begin{equation}\label{relationas}
6.00\frac{m_q \beta_{\pi}}{f_{\pi}^2}\cong
1.12\left(\frac{m_q}{\beta_{\pi}}+1.31\right)
\left(\frac{m_q}{\beta_{\pi}}+5.47 \times 10^1\right).
\end{equation}

There are two basic type of contributions to $F_{\pi\gamma}(Q^2)$
\cite{bhl,huangwu}, i.e.
\begin{equation}\label{ffbas}
F_{\pi\gamma}(Q^2)=F^{(V)}_{\pi \gamma}(Q^2)+F^{(NV)}_{\pi
\gamma}(Q^2).
\end{equation}
$F^{(V)}_{\pi\gamma}(Q^2)$ involves the direct annihilation of
$(q\bar{q})$-pair into two photons, which is the leading Fock-state
contribution that dominates the large $Q^2$ contribution.
$F^{(NV)}_{\pi\gamma}(Q^2)$ involves the case of one photon coupling
`inside' the LC wavefunction of $\pi$ meson , i.e. strong
interactions occur between the photon interactions that is related
to the higher Fock states' contributions.

By keeping the transverse-momentum dependence in both the hard
scattering amplitude and the LC wavefunction, the valence quark
state transition form factor $F^{(V)}_{\pi\gamma}(Q^2)$ can be
written as
\begin{equation}
F^{(V)}_{\pi\gamma}(Q^2) = 2 \sqrt{3}e_\pi\int_0^1 [dx]\int
\frac{{\rm d}^2 \mathbf{k}_\perp}{16 \pi^3} \Psi_{\pi}(x,{\bf
k}_{\perp}) T_H(x,x^{\prime},\mathbf{k}_\perp)
\end{equation}
where $[dx]=dxdx'\delta(1-x-x')$, $e_\pi=(e_u^2-e_d^2)$ and the
hard-scattering amplitude $T_H(x,x',\mathbf{k}_\perp)$ takes the
form
\begin{displaymath}
T_H(x,x',\mathbf{k}_\perp)= \frac{\mathbf{q}_\perp \cdot (x'
\mathbf{q}_\perp + \mathbf{k}_\perp)} {\mathbf{q}_\perp^2 (x'
\mathbf{q}_\perp + \mathbf{k}_\perp)^2} +(x \leftrightarrow x').
\end{displaymath}
$F^{(V)}_{\pi\gamma}(Q^2)$ can be further simplified as the model
wavefunction depends on $\mathbf{k}_\perp$ through $k_\perp^2$ only,
i.e. $\Psi_{\pi}(x,\mathbf{k}_\perp)=\Psi_{\pi}(x,k_\perp^2)$,
\begin{equation}\label{simplea}
F^{(V)}_{\pi \gamma}(Q^2) = \frac{\sqrt{3}e_{\pi}}{4\pi^2}
\int_0^1\frac{d x}{x Q^2}\int_0^{x^2 Q^2}\Psi_{\pi}(x,k_\perp^2) d
k^2_\perp .
\end{equation}

As for the second type of contribution $F^{(NV)}_{\pi\gamma}(Q^2)$,
it is difficult to be calculated in any $Q^2$ region due to its
non-perturbative nature. One can construct a phenomenological model
for $F^{(NV)}_{\pi\gamma}(Q^2)$ based on the asymptotic behavior at
$Q^2\to 0$ and $Q^2\to\infty$. As suggested in Ref.\cite{huangwu},
we assume it takes the following form:
\begin{equation} \label{model}
F^{(NV)}_{\pi\gamma}(Q^2)=\frac{\alpha}{(1+Q^2/\kappa^2)^2},
\end{equation}
where $\kappa$ and $\alpha$ are two parameters that are determined
by the asymptotic behaviors at $Q^2\to 0$, i.e.
\begin{equation}
\alpha=\frac{1}{2}F_{\pi\gamma}(Q^2)|_{Q^2\to 0}\;\;{\rm and}\;\;
\kappa=\sqrt{-\frac{2\alpha}{\frac{\partial}{\partial
Q^2}F^{(NV)}_{\pi \gamma}(Q^2)|_{Q^2\to 0}}},
\end{equation}
where the first derivative of $F^{(NV)}_{\pi \gamma}(Q^2)$ over
$Q^2$ takes the form
\begin{displaymath}
F^{(NV)'}_{\pi\gamma}(Q^2)|_{Q^2\to 0}=
\frac{\sqrt{3}e_{\pi}}{8\pi^2}\left[\frac{\partial}{\partial
Q^2}\int_0^1\int_{0}^{x^2 Q^2}\left(\frac{\Psi_{\pi}(x,k_\perp^2)}
{x^2 Q^2}\right)dx dk_\perp^2\right]_{Q^2\to 0}.
\end{displaymath}
From the phenomenological formula (\ref{model}), it is easy to find
that $F^{(NV)}_{\pi\gamma}(Q^2)$ will be suppressed by $1/Q^2$ to
$F^{(V)}_{\pi \gamma}(Q^2)$ in the limit $Q^2\to\infty$.

\subsection{$\eta\gamma$ and $\eta'\gamma$ Transition form factors}

As for the LC wavefunctions for the pseudoscalars
$\eta_q=\frac{1}{\sqrt{2}} |u\bar{u}+d\bar{d}\rangle$ and
$\eta_s=|s\bar{s}\rangle$, they can be modeled as \cite{hms}
\begin{eqnarray}
\Psi_{\eta_i}(x,\mathbf{k}_\perp)&=&A_{i}\left[\exp\left(-\frac{{\bf
k}_{\perp}^2 +m_i^2}{8{\beta_{i}}^2x(1-x)}\right)\chi^K(m_i,x,{\bf
k}_{\perp})\right] ,
\end{eqnarray}
where $i=q$, $s$ respectively. They also depends on
$\mathbf{k}_\perp$ through $k_\perp^2$ only, i.e.
$\Psi_{\eta_i}(x,\mathbf{k}_\perp)=\Psi_{\eta_i}(x,k_\perp^2)$.
Substituting them into the normalization (\ref{decayconstant}), we
obtain
\begin{equation}
\label{finalp} \int_0^1 \frac{A_{i} m_i
\beta_i\sqrt{x(1-x)}}{4\sqrt{2}\pi^{3/2}}\left(
\mathrm{Erf}\left[\sqrt{\frac{m_i^2+\mu^2_0}{8\beta_{i}^2
x(1-x)}}\right] -\mathrm{Erf}\left[\sqrt{\frac{m_i^2}{8\beta_{i}^2
x(1-x)}}\right] \right)dx=\frac{f_{i}}{2\sqrt{3}},
\end{equation}
where $\mu_0$ stands for the factorization scale, and following the
discussion in Ref.\cite{huangwu}, we take its value to be
$\mu_0\simeq2$~GeV. Under such choice, one may safely set
$\mu_0\to\infty$ to simplify the computation, e.g.
$\mathrm{Erf}\left[\sqrt{\frac{m_i^2+\mu^2_0}{8\beta_{i}^2
x(1-x)}}\right]|_{\mu_0\to\infty}\to 1$, due to the fact that the
contribution from higher $|\mathbf{k}_\perp|$ region to the
wavefunction normalization drops down exponentially for the above
model wavefunctions.

Under the quark-flavor mixing scheme, the $\eta\gamma$ and
$\eta'\gamma$ transition form factors take the following forms:
\begin{eqnarray}
F_{\eta\gamma}(Q^2)&=&F_{\eta_q\gamma}(Q^2)\cos\phi
-F_{\eta_s\gamma}(Q^2)\sin\phi \label{ffeta}
\end{eqnarray}
and
\begin{eqnarray}
F_{\eta'\gamma}(Q^2)&=&F_{\eta_q\gamma}(Q^2)\sin\phi
+F_{\eta_s\gamma}(Q^2)\cos\phi \ ,\label{ffetap}
\end{eqnarray}
where $F_{\eta_q\gamma}(Q^2)$ and $F_{\eta_s\gamma}(Q^2)$ stand for
the $\eta_q\gamma$ and $\eta_s\gamma$ form factors respectively.
Similar to the pion-photon transition form factor, the pseudoscalar
form factors $F_{P \gamma}(Q^2)$ ($P=\eta_q$ and $\eta_s$) can also
be divided into the following two parts,
\begin{equation}\label{ffbasica}
F_{P \gamma}(Q^2)=F^{(V)}_{P \gamma}(Q^2)+F^{(NV)}_{P
\gamma}(Q^2).
\end{equation}
The leading Fock-state contribution $F^{(V)}_{P \gamma}(Q^2)$ can be
simplified as Eq.(\ref{simplea}), and we only need to replace
$e_\pi$ and $\Psi_\pi$ there to the present case of $e_P$ and
$\Psi_P$, where $e_P=(e_u^2+e_d^2,\; \sqrt{2}e_s^2)$ for
$P=(\eta_q,\;\eta_s)$ respectively. And similar to Eq.(\ref{model}),
we assume the following form for the power suppressed non-leading
Fock-state contribution $F^{(NV)}_{P \gamma}(Q^2)$:
\begin{equation} \label{modela}
F^{(NV)}_{P \gamma}(Q^2)=\frac{\alpha}{(1+Q^2/\kappa^2)^2},
\end{equation}
where $\kappa$ and $\alpha$ are two parameters that are determined
by:
\begin{equation}
\alpha=\frac{1}{2}F_{P\gamma}(Q^2)|_{Q^2\to 0}\;\;{\rm and}\;\;
\kappa=\sqrt{-\frac{2\alpha}{\frac{\partial}{\partial
Q^2}F^{(NV)}_{P \gamma}(Q^2)|_{Q^2\to 0}}},
\end{equation}
where the first derivative of $F^{(NV)}_{P \gamma}(Q^2)$ over $Q^2$
takes the form
\begin{displaymath}
F^{(NV)'}_{P \gamma}(Q^2)|_{Q^2\to 0}=
\frac{\sqrt{3}e_P}{8\pi^2}\left[\frac{\partial}{\partial
Q^2}\int_0^1\int_{0}^{x^2 Q^2}\left(\frac{\Psi_{P}(x,k_\perp^2)}
{x^2 Q^2}\right)dx dk_\perp^2\right]_{Q^2\to 0}.
\end{displaymath}

Naively, under strict $SU(3)_F$ symmetry, one has
$F_{\eta_q\gamma}(Q^2)\cong F_{\eta_s\gamma}(Q^2)$, which leads to
\begin{equation}
F_{\eta\gamma}(Q^2)\propto \cos(\phi+45^{\circ})\;\;{\rm and}\;\;
F_{\eta'\gamma}(Q^2)\propto \sin(\phi+45^{\circ})\; .
\end{equation}
Therefore, if $\phi$ varies within the region of
$[0^{\circ},45^{\circ}]$ that is most probably the case,
$F_{\eta\gamma}(Q^2)$ will decrease with the increment of $\phi$,
while $F_{\eta'\gamma}(Q^2)$ will increase with the increment of
$\phi$. In the next section, we will show that under the case of the
broken $SU(3)_F$ symmetry, such a fact is still exist. And then a
possible range of $\phi$ can be obtained by comparing with the
experimental data on the transition form factors
$F_{\eta\gamma}(Q^2)$ and $F_{\eta'\gamma}(Q^2)$.

\section{Numerical analysis}

With the help of the constraints from two photon decay amplitudes of
$\eta$ and $\eta'$ (e.g. Eqs.(\ref{decayfq},\ref{decayfs})) and the
experimental data on the $\eta\gamma$ and $\eta'\gamma$ transition
form factors \cite{CLEO,CELLO,tpc,l3}, one can obtain a reasonable
region for $\phi$. There are several parameters in the wavefunction
$\Psi_P$ to be determined. As for the constitute quark masses, we
take the conventional values: $m_{u,\; d}=300$~MeV and
$m_s=450$~MeV. By studying the pion-photon transition form factor,
one may observe that the best fit of the experimental data is
derived under the case of $m_{u,\; d}\simeq300$~MeV \cite{huangwu}.
For the transverse parameters $\beta_{\pi}$, $\beta_q$ and $\beta_s$
they are proportional to the inverse of the charged radius of the
corresponding valence quark states \cite{huangwu2}. The difference
between them are less than $\sim 10\%$ as shown in Ref.\cite{choi}.
For simplicity, we assume $\beta_q=\beta_s=\beta_{\pi}$ throughout
this work. Uncertainties from the different choices of $\beta_q$,
$\beta_s$, $m_{u,\; d}$ and $m_s$ will be discussed in Sec.IV.

\begin{figure}
\centering
\includegraphics[width=0.45\textwidth]{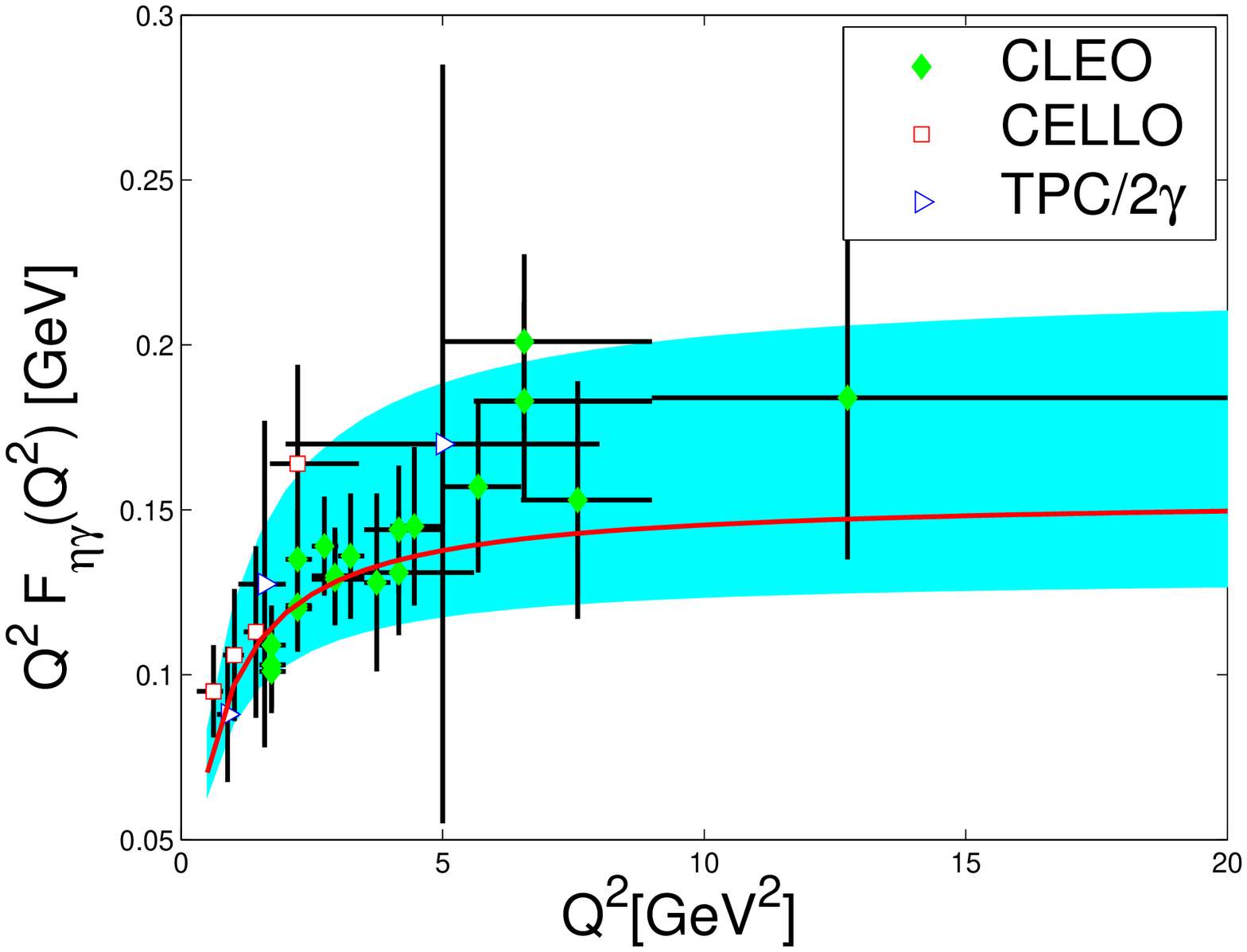}
\hspace{0.1cm}
\includegraphics[width=0.45\textwidth]{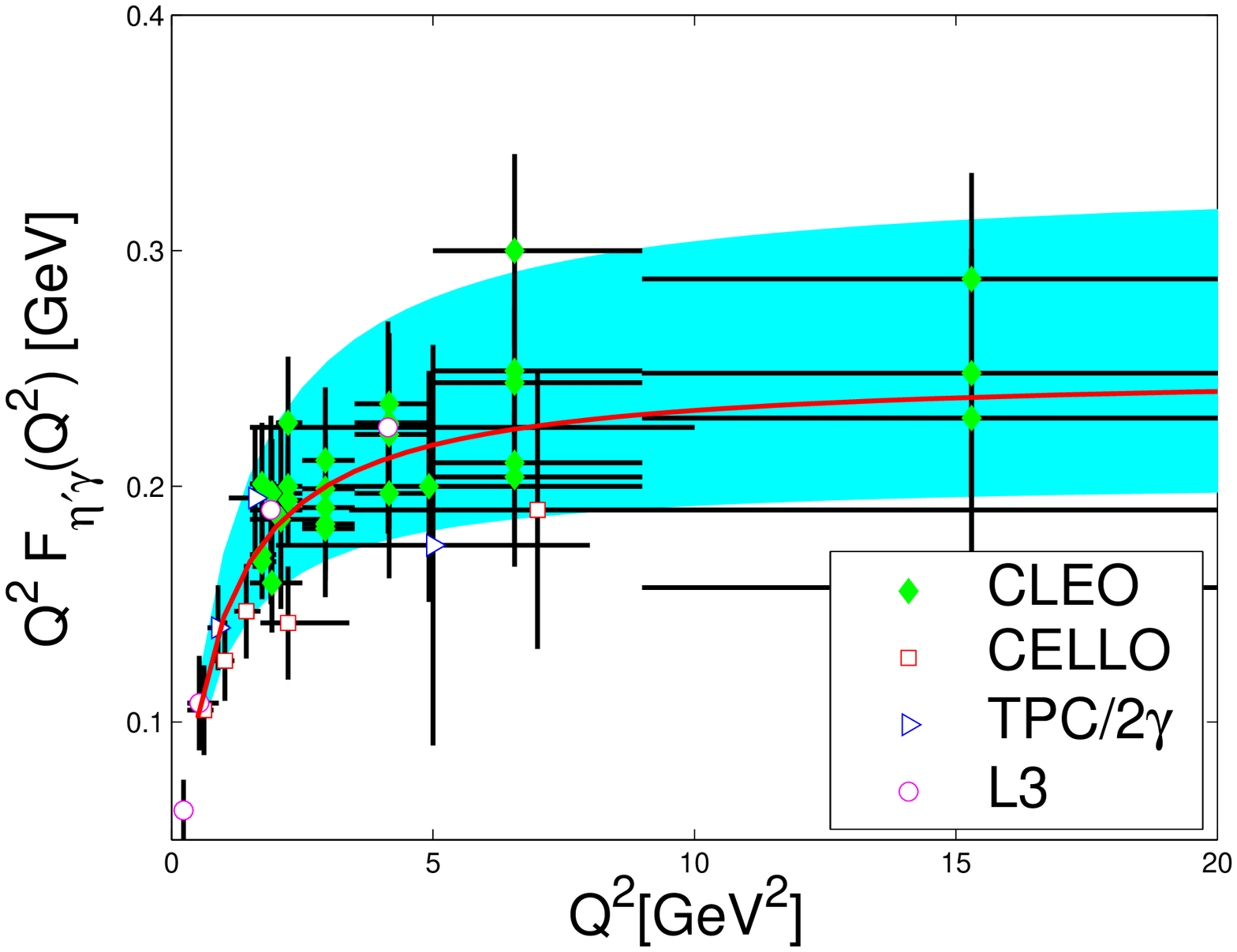}
\caption{Color on-line: Pole-mass fit of $Q^2F_{\eta\gamma}(Q^2)$
(Left) and $Q^2F_{\eta'\gamma}(Q^2)$ (Right) from the experimental
data \cite{CLEO,CELLO,tpc,l3}. The solid line stands for the average
pole-mass fit with $\bar\Lambda_{\eta}=771$~MeV or
$\bar\Lambda_{\eta'}=850$~MeV and the shaded band shows the
experimental uncertainty.} \label{etaexp}
\end{figure}

As for the experimental results of $F_{\eta\gamma}(Q^2)$ and
$F_{\eta'\gamma}(Q^2)$, we take the pole-mass parameter fit formula
that is adopted in those experiments \cite{CLEO,CELLO,tpc,l3}
\begin{equation}\label{polemass}
Q^2F_{(\eta/\eta')\gamma}(Q^2)=\frac{1}{(4\pi\alpha)^2}
\sqrt{\frac{64\pi\Gamma\left[(\eta/\eta')\to\gamma\gamma\right]}
{M_P^3}}\frac{Q^2}{1+Q^2/\Lambda_P^2} .
\end{equation}
As for the values of $\Lambda_{\eta}$ and $\Lambda_{\eta'}$:
\begin{eqnarray}
\Lambda_{\eta}=774\pm11\pm16\pm 22\;\;{\rm MeV} \;\;{\rm and}\;\;
\Lambda_{\eta'}=859\pm9\pm18\pm20\;\;{\rm MeV}
\end{eqnarray}
for CLEO collaboration \cite{CLEO};
\begin{eqnarray}
\Lambda_{\eta}=0.70\pm0.08\;\;{\rm GeV} \;\;{\rm and}\;\;
\Lambda_{\eta'}=0.85\pm0.07\;\;{\rm GeV}
\end{eqnarray}
for TPC/Two-Gamma collaboration \cite{tpc};
\begin{equation}
\Lambda_{\eta'}=900\pm46\pm22\;\;{\rm MeV}
\end{equation}
for L3 collaboration \cite{l3}; and
\begin{eqnarray}
\Lambda_{\eta}=0.84\pm0.06\;\;{\rm GeV} \;\;{\rm and}\;\;
\Lambda_{\eta'}=0.79\pm0.04\;\;{\rm GeV}
\end{eqnarray}
for CELLO collaboration \cite{CELLO}. Averaging the above
experimental values, we obtain the center value for $\Lambda_{\eta}$
and $\Lambda_{\eta'}$, i.e. $\bar\Lambda_{\eta}=771$~MeV and
$\bar\Lambda_{\eta'}=850$~MeV. We draw the pole-mass fit of the form
factors $Q^2F_{\eta\gamma}(Q^2)$ and $Q^2F_{\eta'\gamma}(Q^2)$ in
Fig.(\ref{etaexp}), where the shaded band is derived by adopting the
pole-mass fit formula (\ref{polemass}) and by varying
$\Lambda_{\eta}$ and $\Lambda_{\eta'}$ within the widest possible
range allowed by the above experimental results \footnote{There we
do not take the weighted average of these experiments and treat them
on equal footing, as these experiments are concentrate on different
energy regions and only few data are available.}. The shaded band
(region) for $Q^2F_{\eta\gamma}(Q^2)$ and $Q^2F_{\eta'\gamma}(Q^2)$
can be regarded as constraints to determine the $\eta$/$\eta'-$
wavefunctions, i.e. the values of the parameters in the
wavefunctions and also the mixing angle $\phi$ should make
$Q^2F_{\eta\gamma}(Q^2)$ and $Q^2F_{\eta'\gamma}(Q^2)$ within the
region of the shaded bands as shown in Fig.(\ref{etaexp}).

\begin{figure}
\centering
\includegraphics[width=0.45\textwidth]{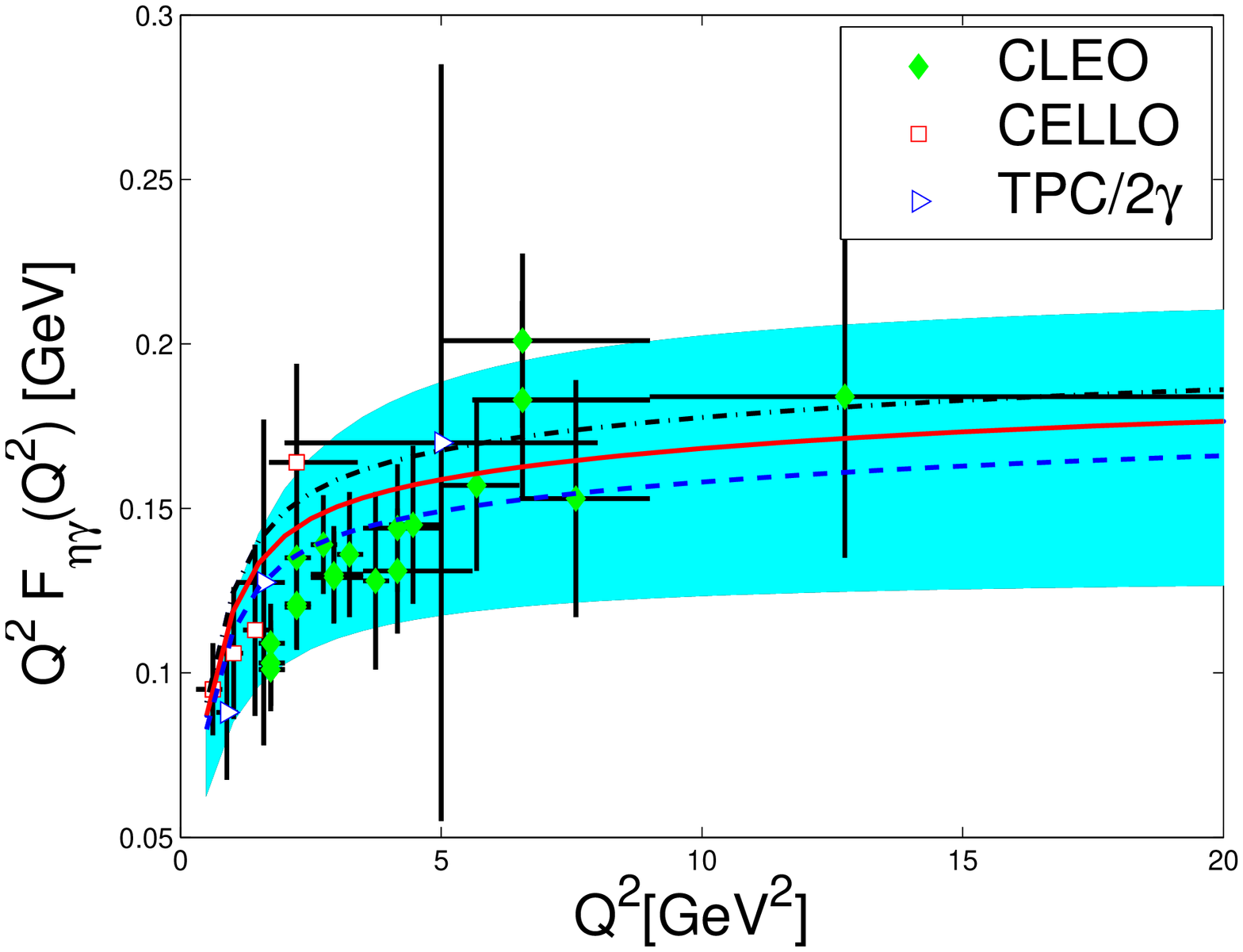}
\hspace{0.1cm}
\includegraphics[width=0.45\textwidth]{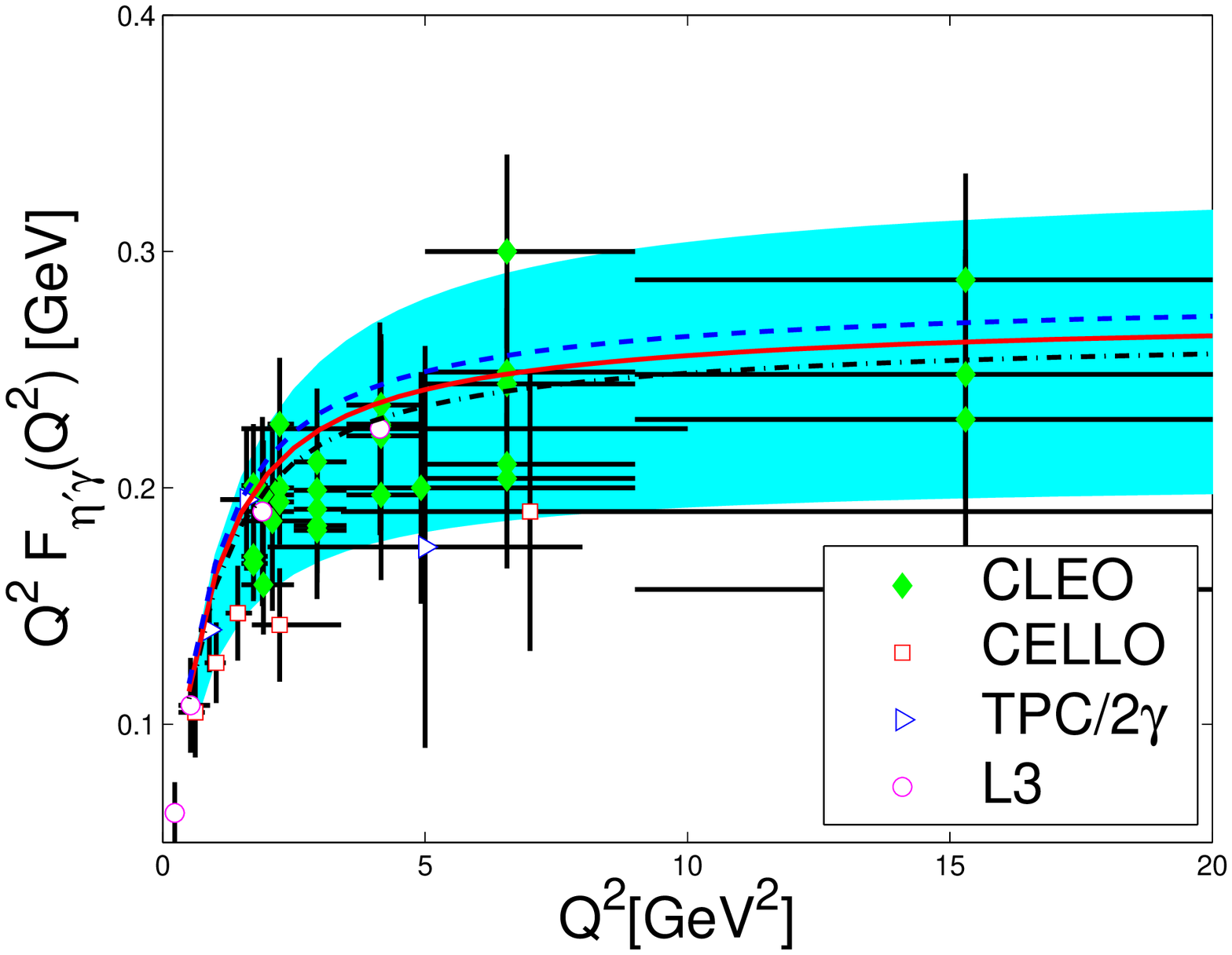}
\caption{Color on-line: $Q^2F_{\eta\gamma}(Q^2)$ and
$Q^2F_{\eta'\gamma}(Q^2)$ with BHL-like wavefunction. The dash-dot
line, the solid line and the dashed line are for $\phi=37.0^\circ$,
$\phi=38.0^\circ$ and $\phi=39.0^\circ$ respectively. It shows that
$Q^2F_{\eta\gamma}(Q^2)$ decreases and $Q^2F_{\eta'\gamma}(Q^2)$
increases with the increment of $\phi$. The shaded band is the
region allowed by the experiments \cite{CLEO,CELLO,tpc,l3}.}
\label{etaexpBHL}
\end{figure}

From Eq.(\ref{finalp}), we obtain
\begin{displaymath}
A_{q}\simeq 2.77\times 10^{2} f_q \;\;\; {\rm and}\;\;\; A_{s}\simeq
2.85\times 10^{2} f_s \ ,
\end{displaymath}
and then with the help of
Eqs.(\ref{decayfq},\ref{decayfs},\ref{ffeta},\ref{ffetap}), it can
be found that only the mixing angle $\phi$ is undetermined. As shown
in Fig.(\ref{etaexpBHL}), one may observe that the value of
$Q^2F_{\eta\gamma}(Q^2)$ decreases with the increment of $\phi$,
while the value of $Q^2F_{\eta'\gamma}(Q^2)$ increases with the
increment of $\phi$, and such a fact can be used to determine a
possible range for $\phi$ by comparing with the experimental data.
In fact, it can be found that the lower limit of $\phi$ is
determined by $Q^2F_{\eta\gamma}(Q^2)$ and the upper limit of $\phi$
is determined by $Q^2F_{\eta'\gamma}(Q^2)$, i.e.
\begin{equation}
\phi\cong38.0^{\circ}\pm 1.0^{\circ} .
\end{equation}
Furthermore, we obtain $\frac{f_q}{f_\pi}=1.07\pm0.01$ and
$\frac{f_s}{f_\pi}=1.24\pm0.10$. By using the correlation between
the quark-flavor mixing scheme and the octet-singlet scheme
\cite{feldmann2,feldmann4}, we obtain
\begin{eqnarray}
&\theta_P=\phi-\arctan\sqrt{2}=-17.0^{\circ}\pm 1.0^{\circ}\;,\\
&\theta_8=\phi - \arctan\frac{\sqrt2 f_s}{f_q}=-20.7^{\circ}\pm
1.0^{\circ}\; ,\theta_1=\phi - \arctan\frac{\sqrt2
f_q}{f_s}=-12.6^{\circ}\pm 3.3^{\circ}
\end{eqnarray}
and
\begin{eqnarray}
&\frac{f_8}{f_\pi}=\sqrt{\frac{f_q^2+2f_s^2}{3}} /
f_\pi=1.19\pm0.07\; ,\frac{f_1}{f_\pi}=
\sqrt{\frac{2f_q^2+f_s^2}{3}}/ f_\pi=1.13\pm0.03 \; ,
\end{eqnarray}

\section{Discussion on the uncertainties of determining $\phi$}

In this section, we discuss the uncertainties for determining the
mixing angle $\phi$ from the above approach. First, we compare the
differences caused by the different model wavefunctions, e.g. the
BHL-like one and the CZ (Chernyak-Zhitnitsky) -like one~\cite{cz}
which is much broad. And then, we restrict ourself to use the
BHL-like model wavefunction for a detail analysis on the effects to
$\phi$ determination caused by each uncertainty sources separately,
where the uncertainty sources mainly include the value of $\beta_q$
and $\beta_s$, the masses of the constitute quarks $u/d$ and $s$,
and the possible intrinsic charm components in $\eta$ and $\eta'$.
Some other even smaller uncertainty sources for the electro-magnetic
transition form factors such as the gluon component in $\eta/\eta'$
will not be discussed \footnote{The gluon contributions might be
important to some other exclusive processes like B meson two-body
non-leptonic exclusive decays. And a discussion on the two gluon
components in the form factors can be found in Ref.\cite{gluonFF}.}.

\subsection{Model dependence}

One typical broad wavefunction is described by the CZ-like
wavefunction. For convenience and simplicity, we take $m_q=0.30$~GeV
and $m_s=0.45$~GeV, and $\beta_{q}=\beta_{s}=\beta_{\pi}$. It can be
found that $\beta_{\pi}=0.70$~GeV for the case of CZ-like
wavefunction \cite{huangwu}.

The CZ-like wavefunctions for $\eta_q$ and $\eta_s$ take the form
\begin{eqnarray}
\Psi^{CZ}_{\eta_i}(x,\mathbf{k}_\perp)&=&A^{CZ}_{i}(1-2x)^2
\left[\exp\left(-\frac{{\bf k}_{\perp}^2
+m_i^2}{8{\beta_{i}}^2x(1-x)}\right)\chi^K(m_i,x,{\bf
k}_{\perp})\right] ,
\end{eqnarray}
where $i=q$, $s$. Following a similar procedure, it can be found
that
\begin{displaymath}
A^{CZ}_{q}\simeq 8.86\times 10^{2} f_q \;\;\; {\rm and}\;\;\;
A^{CZ}_{s}\simeq 8.95\times 10^{2} f_s \ .
\end{displaymath}

Through numerical calculation, it can be found that the $\eta\gamma$
and $\eta'\gamma$ transition form factors from the CZ-like
wavefunction raise faster than the case of the BHL-like
wavefunction. And the shapes of the $\eta\gamma$ and $\eta'\gamma$
form factors from the BHL-like wavefunction are more close to the
pole-mass parameter fits of the experimental data. Although the
model is different, one may find that the value of
$Q^2F_{\eta\gamma}(Q^2)$ decreases with the increment of $\phi$,
while the value of $Q^2F_{\eta'\gamma}(Q^2)$ increases with the
increment of $\phi$. So the range for $\phi$ can also be estimated,
i.e. $\phi\cong 38.0^{\circ}\pm 1.0^{\circ} $, which is close to the
case of BHL-like wavefunction. This shows that the $\eta-\eta'$
mixing angle $\phi$ is almost model-independent in our approach.

Recently, BaBar collaboration has measured the value of
$Q^2F_{\eta\gamma}(Q^2)$ and $Q^2F_{\eta'\gamma}(Q^2)$ at
$Q^2=112\;{\rm GeV}^2$ \cite{babar}:
$Q^2F_{\eta\gamma}(Q^2)=0.229\pm0.030\pm0.008$~GeV and
$Q^2F_{\eta'\gamma}(Q^2)=0.251\pm0.019\pm0.008$~GeV, and the ratio
of the form factors $\kappa=\frac{Q^2F_{\eta'\gamma}(Q^2)}
{Q^2F_{\eta\gamma}(Q^2)}|_{Q^2=112\;{\rm GeV}^2}=1.10\pm0.17$. Under
the case of $\phi\in[37^{\circ},39^{\circ}]$, for the BHL-like
wavefunction, we have $Q^2F_{\eta\gamma}(Q^2)=[0.176,0.190]$~GeV and
$Q^2F_{\eta'\gamma}(Q^2)=[0.228,0.277]$~GeV at $Q^2=112\;{\rm
GeV}^2$, which is close to the experimental values \footnote{It is
also close to the theoretical predictions based on the asymptotic
wavefunction that has been clearly shown in Fig.13 of
Ref.\cite{babar}.} and leads to $\kappa=1.44\pm0.06$. While for the
CZ-like wavefunction, we have
$Q^2F_{\eta\gamma}(Q^2)=[0.267,0.297]$~GeV and
$Q^2F_{\eta'\gamma}(Q^2)=[0.388,0.411]$~GeV at $Q^2=112\;{\rm
GeV}^2$, which is somewhat bigger than the experimental values and
leads to $\kappa=1.41\pm0.06$. As a comparison, one may conclude
that the asymptotic behavior of the form factors
$Q^2F_{\eta\gamma}(Q^2)$ and $Q^2F_{\eta'\gamma}(Q^2)$ disfavor the
CZ-like wavefunction but favor the asymptotic like wavefunction.

It is due to these differences that the form factors of CZ-like
wavefunction and the BHL-like one are affected differently by the
following considered uncertainty sources. For example, as shown in
Ref.\cite{huangwu}, the best fit of $\pi\gamma$ form factor to the
experimental data is obtained with $m_q\simeq 300$ MeV in the case
of BHL-like wavefunction, which is shifted to $m_q\simeq 400$ MeV
for the case of CZ-like wavefunction.

In the following, we will only take the BHL-like wavefunction as an
explicit example to show the uncertainties and the case of CZ-like
one can be done in the similar way. For clarity, in studying of the
uncertainty caused by a certain source, the other uncertainty
sources are taken to be their center values as adopted above.

\subsection{Uncertainty $\Delta\phi^{m}$ from $m_q$ and $m_s$}

We take a wider range for $m_q$ and $m_s$ to study their effects to
the mixing angle $\phi$, e.g. $m_q=0.30\pm0.10$~GeV and
$m_s=0.45\pm0.10$~GeV. Under the present case we adopt
$\beta_p=\beta_s=\beta_{\pi}$, where the value of $\beta_{\pi}$
varies within the region of $[0.48,0.70]\;{\rm GeV}$ according to
the value of $m_q$ \cite{huangwu}. From Eq.(\ref{finalp}), we obtain
the uncertainty from the constituent quark masses
\begin{displaymath}
A_{p}\simeq 2.77^{+1.00}_{-0.48}\times 10^{2} f_p \;\;\; {\rm
and}\;\;\; A_{s}\simeq 2.85^{+1.99}_{-1.10}\times 10^{2} f_s \ ,
\end{displaymath}
where both $A_p$ and $A_s$ increase with the increment of $m_q$ and
$m_s$ respectively. And, it can be found numerically that
\begin{equation}
\Delta\phi^m \leq \pm 0.5^{\circ}.
\end{equation}

\subsection{Uncertainty $\Delta\phi^{\beta}$ from $\beta_q$ and $\beta_s$}

Due to SU(3)-symmetry breaking, there are differences among
$\beta_{\pi}$, $\beta_q$ and $\beta_s$, which is smaller than $10\%$
through a light-cone quark model analysis \cite{choi}. For clarity,
we choose broader ranges $\beta_q=0.55\pm 0.10$~GeV and
$\beta_s=0.55\pm 0.10$~GeV to make a discussion on how these
transverse size parameters affect the mixing angle. Other
wavefunction parameters are fixed by setting $m_q=0.30$~GeV (or
equivalently $\beta_{\pi}=0.55$~GeV) and $m_s=0.45$~GeV. Under such
condition, we have the uncertainty from the transverse parameters
$\beta_q$ and $\beta_s$,
\begin{displaymath}
A_{p}\simeq 2.77^{+1.31}_{-0.69}\times 10^{2} f_p \;\;\; {\rm
and}\;\;\; A_{s}\simeq 2.85^{+1.94}_{-0.89}\times 10^{2} f_s \ ,
\end{displaymath}
where both $A_p$ and $A_s$ increase with the decrement of $\beta_p$
and $\beta_s$, respectively. And, it can be found numerically that
\begin{equation}
\Delta\phi^\beta \leq \pm 2.0^{\circ}.
\end{equation}

\subsection{Uncertainty $\Delta\phi^{c}$ from the intrinsic charm
component }

It has been suggested that a larger intrinsic charm component might
be possible to explain the abnormally large production of $\eta'$ in
the standard model \cite{bpi}. However, some studies in the
literature disfavor such a large portion of intrinsic charm
component, e.g. \cite{huangcao,feldmann,yeh,fc1} and references
therein. It is has been found \cite{feldmann} that the mixing
between the $c\bar{c}$ state with $q\bar{q}$-$s\bar{s}$ basis is
quite small, e.g. less than $2\%$. So for simplicity, we do not
consider the mixing between $c\bar{c}$ and $q\bar{q}$-$s\bar{s}$
basis. And then, we have
\begin{eqnarray}
F_{\eta\gamma}(Q^2)&=&F_{\eta_q\gamma}(Q^2)\cos\phi
-F_{\eta_s\gamma}(Q^2)\sin\phi +F^{\eta}_{\eta_c\gamma}(Q^2)\\
F_{\eta'\gamma}(Q^2)&=&F_{\eta_q\gamma}(Q^2)\sin\phi
+F_{\eta_s\gamma}(Q^2)\cos\phi +F^{\eta'}_{\eta_c\gamma}(Q^2),
\end{eqnarray}
where $F^{\eta}_{\eta_c\gamma}(Q^2)$ and
$F^{\eta'}_{\eta_c\gamma}(Q^2)$ corresponds to the contributions
from the intrinsic charm component in $\eta$ and $\eta'$
respectively, which will be calculated in the following.

The wavefunction of the ``intrinsic" charm component
$\eta_c=|c\bar{c}\rangle$ can be modeled as
\begin{eqnarray}
\Psi^c_{\eta/\eta'}(x,\mathbf{k}_\perp)&=&A^{c}_{\eta/\eta'}
\left[\exp\left(-\frac{{\bf k}_{\perp}^2
+m_c^2}{8{\beta_{c}}^2x(1-x)}\right)\chi^K(m_c,x, {\bf
k}_{\perp})\right],
\end{eqnarray}
where we adopt $\beta_c=\beta_{\pi}=0.55$~GeV and $m_c=1.5$~GeV
\footnote{By varying $\beta_c$ and $m_c$ within their possible
regions, the following results will be slightly changed and the
present conceptional results are the same.}. The overall factor
$A^{c}_{\eta/\eta'}$ is determined by the wavefunction normalization
similar to Eq.(\ref{finalp}), which shows
\begin{displaymath}
A^{c}_\eta=9.45\times10^{3}f^c_{\eta} \;\;\;{\rm and}\;\;\;
A^{c}_{\eta'}=9.45\times10^{3}f^c_{\eta'}\ ,
\end{displaymath}
where $f^c_\eta$ and $f^c_{\eta'}$ are related through
\cite{feldmann},
\begin{equation}
\frac{f^c_\eta}{f^c_{\eta'}}=-\tan \left[\phi - \arctan\frac{\sqrt2
f_s}{f_q}\right]
\end{equation}
and if taking $\phi= 38.0^{\circ}\pm 1.0^{\circ}$, we have
$\frac{f^c_\eta}{f^c_{\eta'}}=[0.36,0.40]$.

It is noted that different from the $\eta_q\gamma$ and
$\eta_s\gamma$ transition form factors, the helicity-flip amplitude
that is proportional to the current quark mass cannot be ignored for
the present case. For the $\eta_c \gamma$ transition form factor, a
direct calculation shows \footnote{There we will not consider the
non-valence charm quark states' contribution since it is quite small
due to the large charm mass effect.}
\begin{eqnarray}
F^{\eta/\eta'}_{\eta_c \gamma}(Q^2) &=& 2\sqrt{6}e_c^2 \int_0^1
[dx]\int \frac{{\rm d}^2 \mathbf{k}_\perp}{16 \pi^3}
\Psi^{c}_{\eta/\eta'}(x,{\bf k}_{\perp})
T^c_H(x,x^{\prime},\mathbf{k}_\perp).
\end{eqnarray}
where the hard-scattering amplitude $T^c_H(x,x',\mathbf{k}_\perp)$
that includes all the helicity states
$(\lambda_1+\lambda_2=0\;,\pm1)$ of $\eta_c$ takes the form
\cite{huangcao,huangcao2},
\begin{equation}
T^c_H(x,x',\mathbf{k}_\perp)= \frac{\mathbf{q}_\perp \cdot (x'
\mathbf{q}_\perp + \mathbf{k}_\perp)} {\mathbf{q}_\perp^2 [(x'
\mathbf{q}_\perp + \mathbf{k}_\perp)^2+m_c^2]} +(x \leftrightarrow
x').
\end{equation}
The above formula can be further simplified by doing the integration
over the azimuth angle:
\begin{equation}
F_{\eta_c \gamma}(Q^2) = \frac{\sqrt{6} e_c^2}{4\pi^2}
\int_0^1\frac{d x}{x
Q^2}\int_0^{\infty}\Psi^{c}_{\eta/\eta'}(x,k_\perp^2)
\left[1+\frac{1-z-y^2}{\sqrt{(z+(1-y)^2)(z+(1+y)^2)}}\right] k_\perp
d k_\perp ,
\end{equation}
where $z=\frac{m_c^2}{x^2 Q^2}$ and $y=\frac{k_\perp}{xQ}$.

\begin{figure}
\centering
\includegraphics[width=0.45\textwidth]{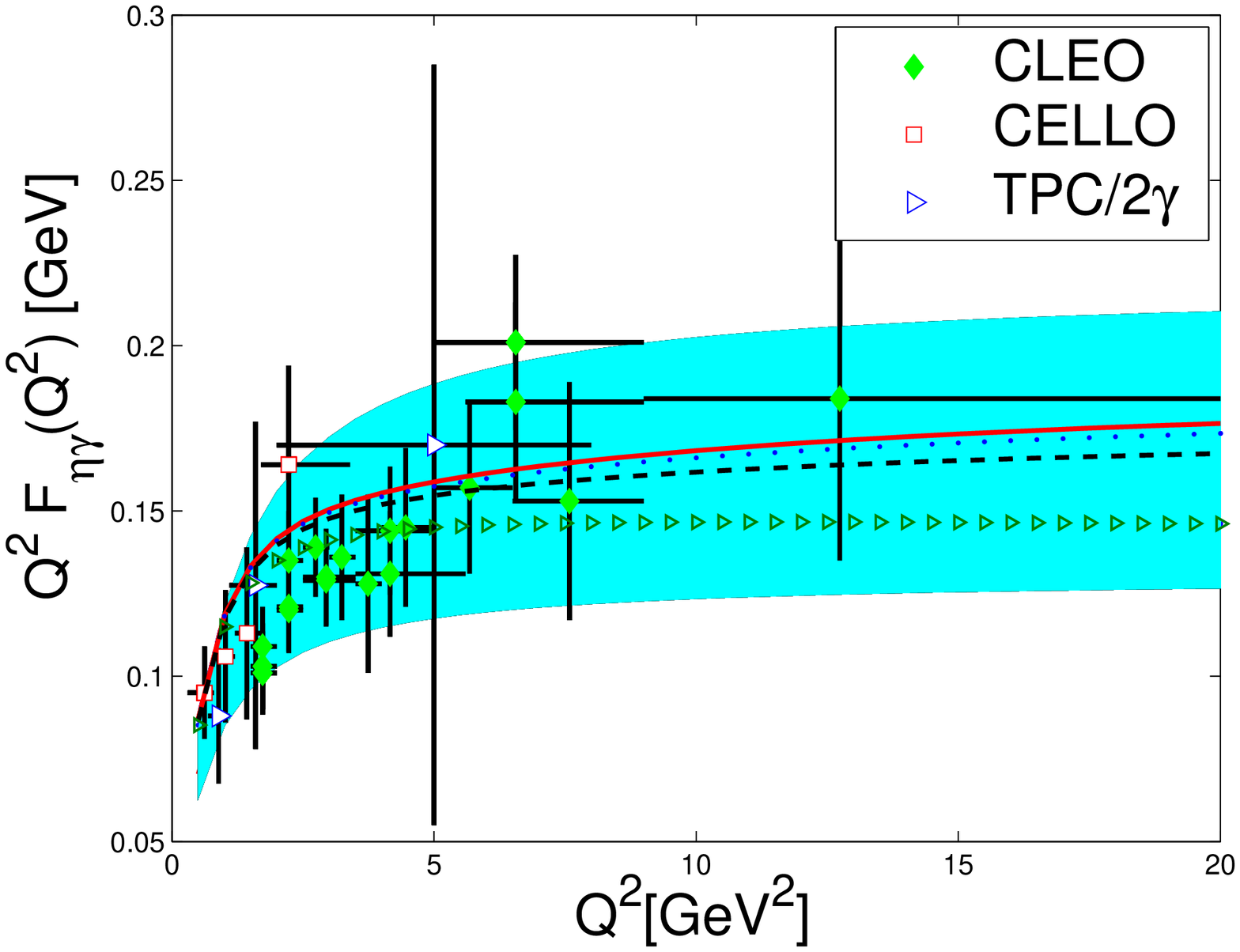}
\hspace{0.1cm}
\includegraphics[width=0.43\textwidth]{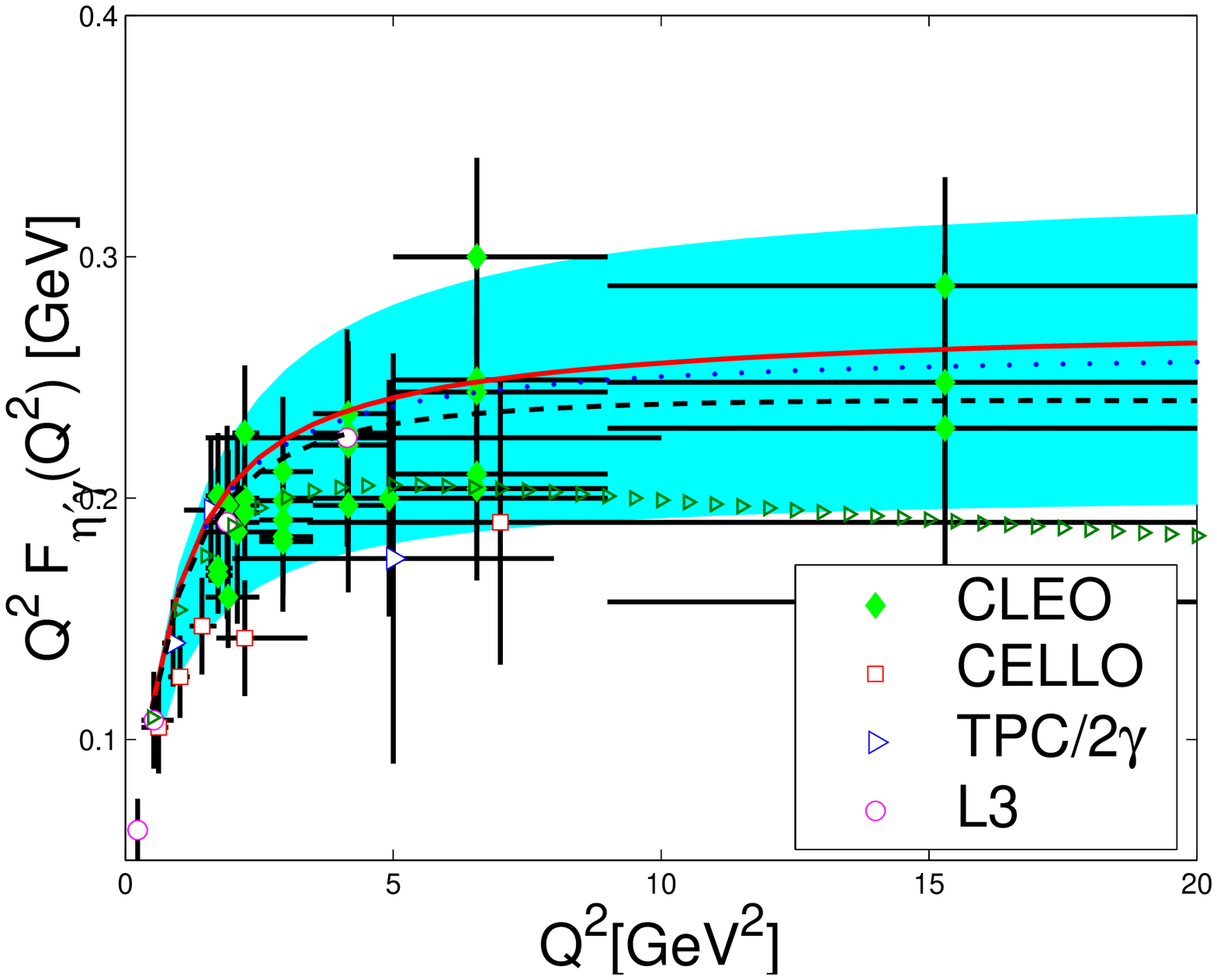}
\caption{$Q^2F_{\eta\gamma}(Q^2)$ and $Q^2F_{\eta'\gamma}(Q^2)$ for
$\phi=38.0^{\circ}$ with BHL-like wavefunction. The shaded band
shows the experimental uncertainty \cite{CLEO,CELLO,tpc,l3}. The
dash-dot line, the solid line, the dashed line and the triangle line
are for $f^c_{\eta'}=0$, $-5$ MeV, $-15$ MeV and $-50$ MeV
respectively.} \label{charm}
\end{figure}

Taking $\phi=38.0^{\circ}$, we draw in Fig.(\ref{charm}) how the
value of $f_{\eta'}^c$ affects the form factors
$Q^2F_{\eta\gamma}(Q^2)$ and $Q^2F_{\eta'\gamma}(Q^2)$. One may
observe that the experimental data disfavors a larger portion of
charm component as $|f_{\eta'}^c|>50$~MeV. Or inversely, it can be
founded that under the condition of $|f_{\eta'}^c|\leq 50$~MeV, the
uncertainty from the possible intrinsic charm components is given by
\begin{equation}
\Delta\phi^{c} \leq \pm 1^{\circ}.
\end{equation}

\subsection{A summary remark on $\phi$}

Under the quark-flavor mixing scheme, and by carefully dealing with
the higher Fock states' contributions, a possible range for $\phi$
can be derived by comparing with the experimental data on the form
factors $F_{\eta \gamma}(Q^2)$ and $F_{\eta' \gamma}(Q^2)$ due to
the fact that the value of $Q^2F_{\eta\gamma}(Q^2)$ decreases, while
the value of $Q^2F_{\eta'\gamma}(Q^2)$ increases, with the increment
of $\phi$ in the whole $Q^2$ region. It has found that the allowable
range for the mixing angle $\phi$ is
\begin{equation}\label{phix1}
\phi\cong38.0^{\circ}\pm 1.0^{\circ} \pm 2.0^{\circ},
\end{equation}
where the first error is from experimental uncertainty
\cite{CLEO,CELLO,tpc,l3} and the second error is from the
uncertainties of the wavefunction parameters and the possible
intrinsic charm component in $\eta$ and $\eta'$. It should be note
that the second uncertainty is lower than the direct sum of the
errors caused by each source separately, i.e.
$\Delta\phi^m+\Delta\phi^\beta+\Delta\phi^c$, which is due to the
fact that these uncertainty sources are correlated to each other.

\section{Summary}

In the present paper, we have performed a light-cone pQCD analysis
of the $\eta\gamma$ and $\eta'\gamma$ transition form factors
$F_{\eta \gamma}(Q^2)$ and $F_{\eta' \gamma}(Q^2)$ involving the
transverse momentum corrections, in which the $\eta-\eta'$ mixing
effects and the contributions beyond the leading Fock state have
been taken into consideration. For such purpose, we have adopted the
quark-flavor mixing scheme for the $\eta$ and $\eta'$ mixing and
have constructed a phenomenological expression to estimate the
contributions beyond the leading Fock state based on its asymptotic
behavior at $Q^2\to 0$ and $Q^2\to\infty$. It has been found that
the value of $Q^2F_{\eta\gamma}(Q^2)$ decreases, while the value of
$Q^2F_{\eta'\gamma}(Q^2)$ increases, with the increment of $\phi$ in
the whole $Q^2$ region, and then a possible range for $\phi$ can be
determined by comparing with the experimental data, which is
$\phi\cong38.0^{\circ}\pm 1.0^{\circ} \pm 2.0^{\circ}$ with the
first error coming from experimental uncertainty
\cite{CLEO,CELLO,tpc,l3} and the second error coming from the
uncertainties of the wavefunction parameters and the possible
intrinsic charm component in $\eta$ and $\eta'$. A more accurate
weighted average of the above mentioned value together with the
seven adopted experimental values as described in
Ref.\cite{feldmann}, and the two new experimental values,
$\phi=41.2^{\circ}\pm1.2^{\circ}$ \cite{kroll} and
$\phi=38.8^{\circ}\pm 1.2^{\circ}$ \cite{bes}, yields
$\bar\phi={39.5^{\circ}} \pm {0.5^{\circ}}$. Furthermore, our
results show that the $\eta-\eta'$ mixing angle $\phi$ depends on
the different wavefunction models slightly. However the asymptotic
behavior of the form factors $Q^2F_{\eta\gamma}(Q^2)$ and
$Q^2F_{\eta'\gamma}(Q^2)$ disfavor the CZ-like wavefunction but
favor the asymptotic like wavefunction. It has been found that the
intrinsic charm component in $\eta$ and $\eta'$ can not be too big,
e.g. $|f^c_{\eta'}|<50$~MeV. Such a conclusion agrees with other
investigations \cite{huangcao,feldmann,yeh,fc1}. These results are
helpful to understand other exclusive processes involving the
pseudo-scales $\eta$ and $\eta'$.

\begin{center}
{\bf ACKNOWLEDGEMENTS}
\end{center}

This work was supported in part by the Natural Science Foundation of
China (NSFC). X.-G. Wu thanks the support from the China
Postdoctoral Science Foundation. \\

\end{document}